# Embodied engagement with scientific concepts: An exploration into emergent learning.

Arlette R.C. Baljon[1], Joseph Alter[2], Marilee Bresciani Ludvik[3]

[1]*Department of Physics, San Diego State University, San Diego, CA*
[2]*Department of Music and Dance, San Diego State University, San Diego, CA*
[3]*Educational Leadership and Policy Studies, University of Texas, Arlington, TX*

**Abstract:** In response to an invitation to integrate science and art pedagogy, science and dance students, enrolled in specific discipline courses, collaboratively produced choreography based on scientific principles. This paper reports students' experience of this process. Science students reported an increased understanding of concepts, while dance students found inspiration for choreography within scientific concepts. Tensions and misconceptions were evident in the process with respect to disciplinary language, the notion of scientific thought, and the level of physical awareness. The relationship between movement and forms of knowledge production in science is investigated as well. The paper ends with recommendations for future classes.

**Introduction**
In 2018 and 2019 two instructors combined their classes on Polymer Science and Dance Making. This was made possible by San Diego State University's (SDSU) Arts Alive initiative, which provides opportunities for students to engage in the arts as an integral part of a comprehensive education that promotes creative research, interdisciplinary collaboration, professional innovation, and personal enrichment. Among the many responses to this initiative was the development of interdisciplinary curriculum that brings faculty and students from arts schools together with faculty and students from academic departments outside of the arts. Teams of faculty provide a transformative educational experience that fosters the understanding of artistic practices as a mode of critical inquiry.

   Interest and focus on such multi-disciplinary work are exponentially increasing fuelled by emerging neuroscience that is illustrating the importance of integrating art disciplines with science disciplines in the hopes that knowledge retention and creative problem solving will abound (Immordino-Yan and Damasio, 2007; Segarra *et al.*, 2018). Given the stereotypes surrounding the learning dispositions and personalities of these oft perceived polar opposite professions, the investigators wondered what it looks like when professors representing each discipline plan a course, deliver it, and then research the outcomes? In this qualitative study, we express our motivation as instructors to engage in this type of interdisciplinary work. This allows the positionality of the investigators to become transparent. The planning and delivery process for this experience is then described in detail, sharing the findings from the student focus groups within the design context. Finally, we provide recommendations for future courses.

**Motivation of Instructors in Personal Narrative Form**
<u>Arlette Baljon (Physics):</u> *When being told about the Arts-Alive initiative, I was immediately "on board". I have always wondered how we scientists come to understand the laws of universe. I do not believe in the myth of disembodied objectivity* *as a foundation of robust inquiry. Often there is the vision of the brain and then the experiments are started. Such a process assumes that the scientist and science student as active participant. However, science aims to eliminate the human aspects of the theories it accepts. It prefers mechanistic theories of matter, often presented in a mathematical framework. Even though this may be the official scientific language, a deeper look at practices and "ways of knowing" reveals that imagination and bodily knowledge is integral to science (Kounios et al., 2008; Meyers, 2015).*

*Great scientists often have profound sudden and sensual experiences that provide them with insight that led to a breakthrough in their research: so-called "aha-moments" (Zajonc, 2006). McClintock, who received a Nobel Prize for her work on genetic transposition in corn, is often cited claiming "I just know about genes and corn in an internal way and then placed my understanding in the frame of science". This claim is often interpreted as her claiming that she sensed and experienced the corn. Evelyn Fox Keller (1983) writes: "Somehow, she placed herself in their feet". Given such first-person experiences are important for scientific discovery, I was wondering how bodily experiences help students to learn scientific concepts in the classroom. Joseph seemed the perfect partner; after all indigenous people have used dance traditionally to approach and understand nature. The collaboration allowed me to investigate if students obtain a deeper conceptual understanding of the dynamics of polymers by moving their bodies.*

<u>Joseph Alter (Dance):</u> *Dance has traditionally been created and structured through "relationships" to musical or theatrical conventions or, in the case of the Judson Church "post-modernist", through the negation of those relationships. That seemed, to me, extremely limited. I first became interested in using science to inform Dance Studies in 2002 as a graduate student at The Ohio State University. There was tremendous pressure to "justify" dance through the lens of interdisciplinary studies: the prevailing notion was that dance, being a largely non-verbal art-form could not "speak" for itself and therefore must look to other disciplines for the means by which to explain it. This struck me as intuitively incorrect - dance could "speak" for itself if only we could understand more deeply it is language. This led me to study "embodied cognition" as the first step in a journey that has branched into many unexpected places as well as convinced me to radically rethink pedagogy and curriculum.*

*I became interested in investigating how other disciplinary research methodologies and perspectives might influence choreographic processes- investigating alternative perspectives of how natural phenomena self-organize, how systems influence one another, and what emergent properties might arise from their adaptation to choreographic practices. As I have found these investigations very fruitful in my development as an Artist, I felt it was imperative to introduce Dance Majors to this area of inquiry.*

**The Learning Design Process**
Joseph and Arlette started to discuss using scientific concepts to inform dance choreography back in 2008. Their official collaboration started in Fall 2016 when several physics students volunteered to meet with the dance company and explore collectively, which physics concepts, might be able to inspire a dance. They called the dance "Strange Attractors", which in scientific chaos theory stands for "a set of numerical values toward which a system tends to evolve". This dance was well received within the dance community and selected through adjunction for the American College Dance Association Conference Gala Performance. Although the science students immensely enjoyed meeting students from a completely different discipline, questions of how arts can inform science were not addressed. The collaboration was entirely focused on the use of science to inform dance choreography.  However, it helped Arlette and Joseph to become familiar with each other's disciplines and motivated them to submit a proposal for a collaborative teaching project, which was selected by the ARTS Alive competition. They proposed to co-create a science and dance class as an entirely new way to explore scientific concepts- one based on intuition, creativity, and human experience and which learning science supported its efficacy (Zull, 2002; Bresciani Ludvik, 2016).  They didn't suggest abandoning mathematical tools in the teaching of science; they simply hoped to systematically expand the scientist's toolbox with new sensual approaches to connect to nature and the understanding of it. Moreover, they hoped to expand the toolbox of the dancers by showing how scientific content can be used to inform choreography. The relevant learning outcomes are shown in Table 1. To compensate for the design time involved, the instructors received one class release each.

| |
|---|
| an increased curiosity in the field that is not their primary focus. |
| a mind that has the capacity to attend to and consider multiple perspectives |
| the ability to translate and apply ideas from conceptual domains to physical action (from the theoretical to the physical) |
| the ability to translate physical phenomena into conceptual modes (from the physical practice to theoretical models). |
| the ability to successfully engage in group problem solving through cooperation and a shared responsibility and mutual respect for both difference and sameness in their respective artistic and scientific disciplines/culture. |
| the ability to comprehend and articulate in writing, and through movement an understanding of critical thinking in both embodied and conceptual forms. |
| an increased ability to "reframe" questions from multiple perspectives and to promote innovation in their respective domains of knowledge. |
| an enhanced ability to "ideate, generate, and articulate" strategies for communicating knowledge to their peers, in their discipline, and in their communities at large. |
| the ability to investigate, through new perspectives, one's "home" discipline. |

Table 1. Learning Outcomes

In the Spring semesters of 2018 and 2019, Arlette and Joseph taught the "Polymer Science" and "Dance Choreography" collaboratively. Each disciplinary class was taught as usual for approx. 80% of the sessions. The other 20% of class time was spent shared. For the science class it replaced a final presentation based on a literature study. For the dance class the time spent in common served as brainstorm for choreography concepts. A fraction of the students' grades were based on the shared project. The "Polymer Science" class is an elective; the "Dance Choreography" class is required for all majors.  The science students were primarily seniors in chemistry, engineering, and physics; the dance students were freshman and sophomore dance majors. The shared sessions covered nine 50-min class sessions. The first three sessions were devoted to basic exercises in choreography that related relationship and movement and helped students to become familiar with each other. Moreover all students were introduced to science concepts that we judged optimal as basic principles for dance

choreography- such as polymer formation, structure property relations, and memory of polymers. Later in the semester students were spending more and more class time in small interdisciplinary groups of 5-6 to work on their dance choreography. Initially, they were asked to come up with a few possible scientific concepts and how these could inform choreography. This "proposal" was presented to the other groups. After that, each group chose a specific topic and worked on a 5 min presentation. The choreographies were shared with an audience of friends and colleagues. Both science and dance students performed; however we made it very clear that we were looking for a proof of principle not aesthetic fluid motion. The presentations were followed by a discussion with the audience.

**The Study**
The purpose of this qualitative study was to explore how students experienced this intentional design process and to ascertain whether learning emerged. The research questions were:

(1) how did students experience this course?
(2) how well did this course design enhance students' learning?

To aid data collection and analysis, an educational research methodologist was brought in to conduct the focus groups, which took place following the course conclusion. Three 60-minute videotaped focus groups were conducted following approval by San Diego State University. Eight to nine students were present in each of the three focus groups. One focus group consisted of all dance majors, another of all science majors, and a third mixed with both. Of the 25 students total, 5 were Hispanic, and 2 Asian, the others Caucasian. Moreover 14 female and 11 male identifying students attended, however 8 of the male and only 5 of the female students were in the science class. Students ranged in age from 19 to 28.

The methodologist designed the interview protocol following an interview with each instructor. Each instructor's course learning outcomes varied, as did their motivation for engaging in this collaborative design. The methodologist included all of these perspectives into the interview protocol design. In addition to the focus group transcripts, instructors analysed remarks students made about their learning experiences in reflective papers. These remarks were not provided based on instructor prompts. Rather, they emerged in a more authentic way as students shared the experience of their learning when reflecting on what they had learned.

Data analysis of transcribed videos and reflective papers used axial and open-ended coding; leveraging constant comparison so that themes could emerge. While the videos were rich with interesting notations -for example, dancers often acted out their responses to the focus group interview protocol while the science majors sat in their chairs and responded with facts first and then experience- only their transcripts were analysed.

**Findings**
The transcript data analysis resulted in the four findings summarized below. Several statements support each of the findings; these are either from a student enrolled in Polymer Science (science) or Dance Choreography (dance). The first finding relates to the difficulty students had understanding each other across disciplines. The second to the traditional role experiential knowledge plays in artistic and scientific modes of inquiry. The third finding inquires into whether sensual approaches like movement can help students learn about scientific concepts. The fourth informs how scientific perspectives might influence choreographic processes.

*Understanding across disciplines*
Science and dance majors entered this learning experience with widely different assumptions and points of reference. For instance, scientists believe that there is only one correct answer to a question (e.g., one correct representation of nature); after all, most often a test question has only one correct answer. On the other hand, dance students are accustomed to ambiguity and the fact that there are many ways to represent a concept through movement. In the collaborative choreographed experience, science students were focusing on phenomena in the world while dance students on the self-other relation. The students lacked a common language to connect these varying epistemologies.  The instructors had been meeting frequently for several years, during which they slowly build a common vocabulary.  In their course work, both sets of students don't use a common language.  Scientists tend to reason *through* abstract formulas. Dancers, on the other hand, express themselves through movement. The students therefore had a hard time finding the words to fill the gap between the disciplines. They bridged that gap successfully by using visuals.  The quotes shared below illustrate this.

    (dance) I did not realize these fields are so different. We would say, what are you talking about?  But then we found common ground.

    (science) I guess with science words and concepts we do not really understand. It helps that they created also a more visual representation. They came up with the covalent bond like holding hands this way. It helped us understand.

    (dance) I learned how smart dancers are because we have to work collaboratively. Work as a team and figure things out. And I think the fact that there is not always a right and wrong answer; people do not always understand. Like finding that common ground and how to build from that.

    However, students did see the positive side in working through these differences

    (dance) Doing a project like this you have to be open-minded. Try a certain style. We are not that comfortable. It teaches us to be open minded, to be adaptive.

    Moreover, some already were more visual oriented

    (science) I am more like kinematic learner, so I like to use my hands when I learn concepts

*Traditional role of experiential knowledge in dance*
Dancers and scientists approach nature fundamentally different. Dancers seek to experience nature as nature and nature within themselves.  They work iteratively: first they "ask a question"; then they experience themselves in the examination of the question (from a state of equanimity or non-judgment); subsequently they reflect and analyze their experience and re-engaging with the question by introducing a perhaps better question; after which the process repeats. They consider movement a form of non-representational "knowing/discovering": that is non-symbolic. In short, as Joseph Roach (2015) writes in the foreword for "What a Body Can Do": "A body can mind". In their paper on enactive cognition Haoseng *et al.* (2019) discuss a new approach to embodied cognition that is based on the assumption that their exist human cognitive processes that do not make any appeal to internal representational or computational states. This more imaginative mystical experiential definition of "truth" dancers embrace -although not unheard of long ago- is in stark contrast with the scientific "truth" we have come to accept during the last several centuries.  For scientists, nature exists outside them; they practice objectivity, brokerage knowledge and arbitrate the truth. The following statements from students capture this

(science) As a science student, we always learn our scientific concepts thought theory, equations, diagrams, experiments, etc. Sometimes we are lacking imagination on how molecules interact. But for dance students, it is the opposite, they are full of creation, imagination, and express themselves through body movements.

(science) As scientists we believe there is a "truth" in science. We need to explain as best as we can a truth system. This is more my personal opinion; under pressure of time, we were juggling with concepts to portray a truth system. It is hard to stay close to the "truth" if things are simplified. Dancers mix the "truth" with the personal emotional when they interpret the system.

(dance) And there was kind of a barrier because they would say: "this is what it is; how it is supposed to be" and we would say "well we can change things up it is a creative collaborative experiment" From a creative standpoint what can we expand on and what is the truth of the science.

(dance) Being aware of the space we are in as that relates to physics itself. Keeping that lens of scientific and physics concepts in your head while re-questioning yourself. Rather than "I am a bond" "what is this bonding doing in my body". If that makes any sense to you. As a dancer we take our own perspective of it. Like I have a perspective of a polymer and put that into my dancing.

As a result, science students were more focused on the end product (a dance on a scientific concept), while dance students tended to focus on the process of creation itself.

(science) As scientists we need to be consistent. What should we do? What does it mean? It has to be accurate. Has to be expressive. To be able to combine these two opposite perspectives is the challenge. Ok we need an end product, but it does not matter how we get there.

(dance) We make a dance just to remake it. Just to fix it again. Just to say we do not like it. You see what I mean. That is the kind of work we are into. We tried doing things, had setbacks, and tried it again. Not the end product. But look at the steps/pathway.

Other explanations for the difference between the student groups were mentioned as well

(dance) The dance majors were more extroverted and the science majors more reserved. All open to ideas but it was hard to come to action.

*Learning scientific concepts through movement*

Many science students believed that the class was a transformational experience for them. They even mentioned that participating actively in animating forces and relations between molecules helped them to better understand their behaviour.

(science) Before this class, we were more likely to understand the concepts and structures from the description or pictures from the textbook, which means that we are receiving the information passively. However, when we are embodying scientific concepts, we have to express our understanding subjectively, and this would deepen the impression of the knowledge. Besides, embodying the concepts also taught me that there is no wrong way to express your understanding using your body. I used to think objectively, and I thought there would always be one correct answer in the world of science. However, the dance taught me to think subjectively and there is no fixed way to present the concepts, phenomena and structures in our dance.

(science) It was not a traditional lecture style, traditional class time approach. It was more about seeing this. What really struck me is how artistic science can be. Not just what we already know. It is not about learning here that it is more for the lectures; it is more like the creative insight here. You put yourself in the position of the molecules.

(science) Engineering has such a strict foundation for what is considered correct. The boundaries we are allowed to cross are limited as to what is possible for the given environment. Dance making and art has open boundaries and even if crossing them is considered wrong, the end result might turn out to be something beautiful. Think of talented artists. Do they follow the rules of what is conventional? Engineering needs to consider a creative concept while also maintaining the balance of being within technical correctness.

(science) Furthermore, I would say that the embodying of some science concepts can complement the formal study of science. It would not replace it because the math part of the science will still require some calculations and derivations on the paper. Moreover, the problem solving sometimes demands the traditional way of studying. However, the conceptual part of science will one hundred percent benefit from this innovation.

(science) This was the main part of this project that I didn't expect. I didn't expect to come from this experience of creating a dance having any better of an understanding of polymer science than I had already gained from our in-class sessions. But surprisingly, working out how to move our bodies in a way to portray a polymer's behaviour in certain situations actually accelerated my understanding of the topics themselves. It was as if the ideas that I had learned on paper were alive and in front of us. This entire experience has really opened my mind to the many paths of learning there are both in our university and in the world, and to the idea that there is no singular experience when it comes to education. The wonder of learning can come in many different forms, each unique and important in their own ways. The real magic is sharing these experiences with one another and keeping the learning alive.

(science) Lastly, while taking this class as a future chemistry high school teacher I was able to see how the science can be explained to children in the way that all types of learners can understand them. In teaching there is the universal design for learning concept, which states that lessons need to be designed in a way that all students are able to learn. I was struggling with this concept because I did not know how the science could be taught to children who have no interest in it at all, but this class showed me: I need to think outside of the box. Art and science can go hand in hand, and one can be taught through another that way, so all students gain from learning.

However, a few disagreed

(science) I do not think there was any sort of net gain: beneficial or not beneficial for us from teaching the concepts.

(science) I think I have a good understanding of the scientific concepts and movement is not beneficial for reinforcing the concepts. It is difficult to teach a concept with physical movement.

(science) I think the dance is a very avant-garde method of approach. Everything communicated was abstract not concrete.

### *Dance choreography based on scientific concepts*

Although the dance students tried to use the scientific principles to explore and inspire choreographic thought, most science students focused on straightforward translating the scientific concept into choreography. This disconnect was hard to overcome during the limited time they spent together.

(dance) Our dance was showing how bonds form polymers. We just went from single- ourselves- and then made a bond with someone. Just having your arm on them.

(science) So we had this idea to talk about bond formation. And so, when I was trying to explain the different kinds of bonds, I realized that it's more about pockets. So, are we able to show the dancers the difference between these kinds of bonds? So, we actually stood up with them and said, "Okay, so you are going to hold this hand". So maybe we did not talk about electrons and transfer, but they still got the picture.

(dance) Like when I was in a group this is what happened: We started the process, and the scientist would say "all polymers repeat" and then the dancers would think "how can we create a dance from that?" What are the possibilities? And then we would actually create the dance. We wanted it to be accurate to make the science people happy.

(dance) I feel that from my standpoint with the group that we had to take back stuff. Like I set earlier they wanted to do exactly what polymers do. We wanted to take it a little bit further. But we could not do that because of the structure they wanted to follow. When we would have had the collaboration longer, we would have been able to go deeper with that.

(dance) more communication was needed and trying to move out. Experimenting. Everyone has to be on board with this concept. Else you are stuck and cannot go further.

(dance) I feel that from my standpoint with the group that we had to take back stuff. Like I set earlier they wanted to do exactly what polymers do. We wanted to take it a little bit further. But we could not do that because of the structure they wanted to follow. When we would have had the collaboration longer, we would have been able to go deeper with that.

Moreover, science students complained about the background of the dance students

(science) No ours was about how polymers move. We did not do anything scientific besides. The assumption is that they understand atoms. At the level of AP Chem.

(science) We are simple met with a lack of experience. Most parties should have more time invested into each other's fields. So, the scientists should focus more on the dance theory behind it. And within the capabilities of the dancers in this case more instruction on the science aspects. The instructors already selected an umbrella of topics but if we could build on these topics even more. So, for being highly functional for both parties a sort of experience of a type required beyond the 10 meetings together. It is too ambitious.

**Summary and Discussion of Findings**
Dance and science majors participated in innovative courses, which included an assignment to collaboratively choreograph a short dance based on scientific principles to support emergent learning. Despite many obstacles, the majority of students within both groups valued the learning experience. Both groups were driven by a similar curiosity and wish to order information about the world around them. They experienced a friction in how the two disciplines – science and dance- explore the natural world. Scientists construct objective theories extensively relying on mathematics to capture the "truth" of the universe's mysteries (Falk, 2019; Frank *et al*., 2019).   Hence it is no wonder that science students insist that the deepest truths about reality and life can only be revealed through a scientific approach that aims to construct objective knowledge independent of the culture and worldview of the observer.  In this sense, it is evident that the prescriptive learning ecology was at play (Collins and Halverson, 2010).

Dance, on the other hand, serves to connect to the universe in a ritual and ethical way by gathering individual subjective experiential knowledge of nature's mysteries and needs (Evans, 2020).  The complex-adaptive learning ecology is likely at work here (Mitra and Dangwal, 2010).  Both groups of students recognized this divide, and the science students were challenged by the dance students to reconsider these assumptions: first they struggled with and then started to appreciate an embodied approach to learning and robust inquiry. They acknowledged that the artistic approach helped them to understand and remember concepts, as has been suggested in the literature (Immordino-Yang and Damasio, 2017). By looking at science from the perspective of the molecules, they came to appreciate that all "knowing" is embodied. However one can, for the purposes of making distinctions regarding the "qualia of states of knowing" use the terms Conceptual Knowledge (one that employs symbols like language or mathematics) and Embodied Knowledge which has distinct qualia and does not deal with symbolic languages but rather an "ineffable knowing" that language only points to.

Interestingly, the students learning process findings mirror those of Meyers' study of crystallographers' learning process. Meyers (2015) performed ethnographical studies of crystallographers in the field of structural biology and protein modelling and discovered that all of these scholars relied on their moving bodies to reason through the molecular structure of a complex biological molecule and make sense of the molecular realm. Like our students, these scholars learned how to feel through structures by experimenting with the forces and tensions in their own bodies. Unlike the traditional approach in which science is understood as an objective, rational, and disembodied practice, according to Meyers their bodywork seems to be a form of knowing and mode of thought.

Similarly, our findings mirror what we understand to be true about learning. Zull (2002) and Bresciani Ludvik *et al.*, (2016) illustrate that emotion is a part of learning, that movement enhances learning, and that reflection and creativity abound when multiple ways of learning can be introduced, particularly with adolescent learners (ages 10-25). When we consider the role of emotion in an emergent ecology of learning that seeks to combine seemingly disparate ecologies of learning, we recognize that more opportunity for dialogue and exploration must ensue (Williams *et al.*, 2011) so that students can focus on resolving a potential communication and understanding problem (Nilson, 2010).

**Recommendations**

In retrospect, this class design and implementation was very ambitious. The collaboration took place as part of two disciplinary courses and as a result, students spent only approximately 10-15 hours together over the entire semester given required disciplinary course content could not be dropped. Most students mentioned that given the large divide between the disciplines, they needed more time to develop a common framework. This student observation is often echoed in research about emergent learning, where learning ecologies exist and when they come together a new learning ecology must emerge (Williams *et al.*, 2011). One way in which this could be accomplished is through a 3-unit general education classes entirely focused on this interdisciplinary process. We recommend that during the first half of the semester, students are introduced to each other's disciplines and familiarized with each other's disciplinary language and belief systems using a variety of problem-based learning pedagogical practices that appeal to each discipline (Nilson, 2010). Students will benefit from discussions in small interdisciplinary groups as they seek to understand the scientific concepts, how to work with them, and whether there is a problem to be solved in the way that they interact and learn from each other. Changing the composition of the discussion groups will allow them to meet classmates they may not ordinarily meet. In addition, it is important that during this period, the instructors select together a set of teaching materials (both content-related, applied learning-related, and assessment-related) for students to work with in relationship to each other. These can be introduced to the class as a whole.

In the second half of the semester, students are assigned to permanent groups to work on the artistic product. We recommend that they have the opportunity to show and/or perform their product to friends and the campus community at the end of the process. We recommend combining this demonstration of learning with an open discussion about their experiences.

Interdisciplinary collaborative learning that seeks to create an emergent learning ecosystem will likely help both science and non-science students in their future careers and as concerned citizens. We live in a time of increasing distrust towards the scientific community. According to Kabat (2017), trust might be regained when scientists adopt a less authoritarian attitude. That is when they would acknowledge what they do not know instead of presenting a dogmatic truth. Classes like the one we presented here may enable non-science and science students to enter into a collaborative problem-solving dialogue that is intertwined with nature and the human condition as opposed to attempting to remove them. Future scientists would learn how to incorporate embodied experiences to deepen their knowledge and retention of concepts. Art students, on the other hand, would learn to use science effectively as inspiration for artistic experience. Both would hopefully experience a renewed sense of wonder at academic discovery and recognize that creativity and knowledge production are interconnected. After all there can be as much aesthetic beauty in a physical theory as in an artwork.